\documentclass[aoas,MSNbibl,nameyear,dvips]{arximspdf}
\usepackage{multirow,dcolumn,url,breakurl}
\usepackage{graphicx}

\doi{10.1214/13-AOAS683} 
\volume{8}
\issue{2}
\pubyear{2014}
\firstpage{777}
\lastpage{800}

\makeatletter
\newcolumntype{d}[1]{D{.}{.}{#1}}
\newcommand{\rank}{\mathrm{rank}}
\newtheorem{theorem}{Theorem}[section]
\newtheorem{lemma}{Lemma}[section]
\newproclaim{remark}{Remark}
\newproclaim{step}{Step}
\makeatother

\begin{document}
\begin{frontmatter}

\title{Hypothesis setting and order statistic for\\ robust genomic
meta-analysis\thanksref{T1}}
\runtitle{Order statistic for genomic meta-analysis}

\begin{aug}
\author[A]{\fnms{Chi} \snm{Song}\ead[label=e1]{chi.song@yale.edu}}
\and
\author[A]{\fnms{George C.} \snm{Tseng}\corref{}\ead[label=e2]{ctseng@pitt.edu}}
\runauthor{C. Song and G. C. Tseng}
\affiliation{University of Pittsburgh}
\address[A]{Department of Biostatistics\\
Graduate School of Public Health\\
University of Pittsburgh\\
Pittsburgh, Pennsylvania 15261\\
USA\\
\printead{e1}\\
\phantom{E-mail: }\printead*{e2}} 
\end{aug}
\thankstext{T1}{Supported by NIH R21MH094862 and R01DA016750.}

\received{\smonth{6} \syear{2012}}
\revised{\smonth{7} \syear{2013}}

%
\begin{abstract}
Meta-analysis techniques have been widely developed and applied in
genomic applications, especially for combining multiple transcriptomic
studies. In this paper we propose an order statistic of $p$-values ($r$th
ordered $p$-value, rOP) across combined studies as the test statistic. We
illustrate different hypothesis settings that detect gene markers
differentially expressed (DE) ``in all studies,'' ``in the majority of
studies'' or ``in one or more studies,'' and specify rOP as a suitable
method for detecting DE genes ``in the majority of studies.'' We
develop methods to estimate the parameter $r$ in rOP for real
applications. Statistical properties such as its asymptotic behavior
and a one-sided testing correction for detecting markers of concordant
expression changes are explored. Power calculation and simulation show
better performance of rOP compared to classical Fisher's method,
Stouffer's method, minimum $p$-value method and maximum $p$-value method
under the focused hypothesis setting. Theoretically, rOP is found
connected to the na\"{i}ve vote counting method and can be viewed as a
generalized form of vote counting with better statistical properties.
The method is applied to three microarray meta-analysis examples
including major depressive disorder, brain cancer and diabetes. The
results demonstrate rOP as a more generalizable, robust and sensitive
statistical framework to detect disease-related markers.
\end{abstract}

%

\end{frontmatter}

\section{Introduction}\label{sintro}
With the advances in high-throughput experimental technology in the
past decade, the production of genomic data has become affordable and
thus prevalent in biomedical research. Accumulation of experimental
data in the public domain has grown rapidly, particularly of microarray
data for gene expression analysis and single nucleotide polymorphism
(SNP) genotyping data for genome-wide association studies (GWAS). For
example, the Gene Expression Omnibus (GEO; \url{http://www.ncbi.nlm.nih.gov/geo/}) from the National \mbox{Center} for
Biotechnology Information (NCBI) and the Gene Expression Atlas (\url{http://www.ebi.ac.uk/gxa/}) from the European Bioinformatics Institute
(EBI) are the~two largest public depository websites for gene
expression data and the database of Genotypes and Phenotypes (dbGaP,
\url{http://www.ncbi.nlm.nih.gov/gap/}) has the largest collection of
genotype data. Because individual studies usually contain limited
numbers of samples and the reproducibility of genomic studies is
relatively low, the generalizability of their conclusions is often
criticized. Therefore, combining multiple studies to improve
statistical power and to provide validated conclusions has emerged as a
common practice [see recent review papers by \citet{tseng2012comprehensive} and \citet{begum2012comprehensive}]. Such
genomic meta-analysis is particularly useful in microarray analysis and
GWAS. In this paper we focus on microarray meta-analysis while the
proposed methodology can be applied to the traditional ``univariate''
meta-analysis or other types of genomic meta-analysis.

Microarray experiments measure transcriptional activities of thousands
of genes simultaneously. One commonly seen application of microarray
data is to detect differentially expressed (DE) genes 
in samples labeled with
two conditions (e.g., tumor recurrence versus nonrecurrence), multiple
conditions (e.g., multiple tumor subtypes), survival information or
time series. In the literature, microarray meta-analysis usually refers
to combining multiple studies of related hypotheses or \mbox{conditions} to
better detect DE genes (also called candidate biomarkers). 
For this problem, two major types of statistical procedures have been
used: combining effect sizes or combining $p$-values. Generally speaking,
no single method performs uniformly better than the others in all data
sets for various biological objectives, both from a theoretical point
of view [\citeauthor{littell1971asymptotic} (\citeyear{littell1971asymptotic,littell1973asymptotic})] and from
empirical experiences. %
In combining effect sizes, the fixed effects model and the random
effects model are the most popular methods [\citet{cooper2009handbook}]. These methods are usually more straightforward
and powerful to directly synthesize information of the effect size
estimates, compared to $p$-value combination methods. They are, however,
only applicable to samples with two conditions when the effect sizes
can be defined and combined. On the other hand, methods combining
$p$-values provide better flexibility for various outcome conditions as
long as $p$-values can be assessed for integration. Fisher's method is
among the earliest $p$-value methods applied to microarray meta-analysis
[\citet{rhodes2002meta}]. It sums the log-transformed $p$-values
to aggregate statistical significance across studies. Under the null
hypothesis, assuming that the studies are independent and the
hypothesis testing procedure correctly fits the observed data, Fisher's
statistic follows a chi-squared distribution with degrees of freedom
$2K$, where $K$ is the number of studies combined. Other methods such
as Stouffer's method 
[\citet{stouffer1949american}], the minP method [\citet{tippett1931methods}]
and the maxP method [\citet{wilkinson1951statistical}]
have also been widely used in microarray meta-analysis.
It can be shown that these test statistics have simple analytical forms
of null distributions and, thus, they are easy to apply to the genomic
settings. The assumptions and hypothesis settings behind these methods
are, however, very different and have not been carefully considered in
most microarray meta-analysis applications so far. In Fisher, Stouffer
and minP, the methods detect markers that are differentially expressed
in ``one or more'' studies (see the definition of $\mathrm{HS}_B$ in
Section~\ref{shs}). In other words, an extremely small $p$-value in one
study is usually enough to impact the meta-analysis and cause
statistical significance. On the contrary, methods like maxP tend to
detect markers that are differentially expressed in ``all'' studies
(called $\mathrm{HS}_A$ in Section~\ref{shs}) since maxP requires
that all combined $p$-values are small for a marker to be detected. In
this paper, we begin in Section~\ref{shs} to elucidate the hypothesis
settings and biological implications behind these methods. In many
meta-analysis applications, detecting markers differentially expressed
in all studies is more appealing. The requirement of DE in ``all''
studies, however, is too stringent when $K$ is large and in light of
the fact that experimental data are peppered with noisy measurements
from probe design, sample collection, data generation and analysis.
Thus, we describe in Section~\ref{shs} a robust hypothesis setting
(called $\mathrm{HS}_r$) that detects biomarkers differentially
expressed ``in the majority of'' studies (e.g., $>$70\% of the studies)
and we propose a robust order statistic, the $r$th ordered $p$-value
(rOP), for this hypothesis setting.

The remainder of this paper is structured as follows to develop the rOP
method. In Section~\ref{srop} the rationale and algorithm of rOP are
outlined, and the methods for parameter estimation are described in
Section~\ref{sselectr}. Section~\ref{srop1side} extends rOP with a
one-sided test correction to avoid detection of DE genes with
discordant fold change directions across studies. Section~\ref
{sapplication} demonstrates applications of rOP to three examples in
brain cancer, major depressive disorder (MDD) and diabetes, and
compares the result with other classical meta-analysis methods. We
further explore power calculation and asymptotic properties of rOP in
Section~\ref{spowcalc}, and evaluate rOP in genomic settings by
simulation in Section~\ref{ssimu}. We also establish an unexpected
but insightful connection of rOP with the classical na\"ive vote
counting method in Section~\ref{svc}. Section~\ref{sconclusion}
contains final conclusions and discussions.

\section{$r$th ordered $p$-value (rOP)}\label{srOP}

\subsection{Hypothesis settings and motivation}\label{shs}
We consider the situation when $K$ transcriptomic studies are combined
for a meta-analysis where each study contains $G$ genes for information
integration. Denote by $\theta_{gk}$ the underlying true effect size
for gene $g$ and study $k$ ($1\le g\le G$, $1\le k\le K$). For a given
gene $g$, we follow the convention of \citet{birnbaum1954combining} and \citet{li2011adaptively} to consider
two complementary hypothesis settings, depending on the pursuit of
different types of targeted markers:
\begin{eqnarray*}
&& \mathrm{HS}_A\dvtx  \biggl\{H_0\dvtx \bigcap
_k\{\theta_{gk}=0\}\mbox{ versus } H_a^{(A)}\dvtx \bigcap_k\{
\theta_{gk}\ne0\} \biggr\},
\\
&& \mathrm{HS}_B\dvtx  \biggl\{H_0\dvtx \bigcap
_k\{\theta_{gk}=0\}\mbox{ versus } H_a^{(B)}\dvtx \bigcup_k\{
\theta_{gk}\ne0\} \biggr\}.
\end{eqnarray*}
In $\mathrm{HS}_A$, the targeted biomarkers are those differentially
expressed in all studies (i.e., the alternative hypothesis is the
intersection event that effect sizes of all $K$ studies are nonzero),
while $\mathrm{HS}_B$ pursues biomarkers differentially expressed in
one or more studies (the alternative hypothesis is the union event
instead of the intersection in $\mathrm{HS}_A$). Biologically
speaking, $\mathrm{HS}_A$ is more stringent and more desirable to
identify consistent biomarkers across all studies if the studies are
homogeneous. $\mathrm{HS}_B$, however, is useful when heterogeneity is
expected. For example, if studies analyzing different tissues are
combined (e.g., study 1 uses epithelial tissues and study 2 uses blood
samples), it is reasonable to identify tissue-specific biomarkers
detected by $\mathrm{HS}_B$. We note that $\mathrm{HS}_B$ is
identical to the classical union-intersection test (UIT) [\citet{roy1953heuristic}] but $\mathrm{HS}_A$ is different from the
intersection-union test (IUT) [\citet{berger1982multiparameter}, \citet{berger1996bioequivalence}]. In IUT, the
statistical \mbox{hypothesis} is in a complementary form between the null and
alternative hypotheses $ \{H_0\dvtx \bigcup_k\{\theta_{gk}=0\}$
versus $H_a^{(A)}\dvtx \bigcap_k\{\theta_{gk}\ne0\} \}$. Solutions
for IUT require a more sophisticated mixture or Bayesian modeling to
accommodate the composite null hypothesis and are not the focus of this
paper [for more details, see \citet{erickson2009meta}].

As discussed in \citet{tseng2012comprehensive},
most existing genomic meta-analysis methods target on $\mathrm{HS}_B$.
Popular methods include classical Fisher's method [sum of minus
log-transformed $p$-values; \citet{fisher1925statistical}], Stouffer's
method [sum of inverse-normal-transformed $p$-values; \citet{stouffer1949american}],
minP [minimum of combined $p$-values; \citet{tippett1931methods}] and a recently proposed adaptively weighted (AW)
Fisher's method [\citet{li2011adaptively}]. The random effects model
targets on a slight variation of $\mathrm{HS}_A$, where the effect
sizes in the alternative hypothesis are random effects drawn from a
Gaussian distribution centered away from zero (but are not guaranteed
to be all nonzero). The maximum $p$-value method (maxP) is probably the
only method available to specifically target on $\mathrm{HS}_A$ so
far. By taking the maximum of $p$-values from combined studies as the
test statistic, the method requires that all $p$-values be small for a
gene to be detected. Assuming independence across studies and that the
inferences to generate $p$-values in single studies are correctly
specified, $p$-values ($p_k$ as the $p$-value of study $k$) are\vspace*{1.5pt} i.i.d.
uniformly distributed in $[0,1]$. Fisher's statistic
($S^{\mathrm{Fisher}}=-2\sum\log p_k$) follows a\vspace*{1.5pt} chi-squared distribution with
degree of freedom $2K$ [i.e., $S^{\mathrm{Fisher}}\sim\chi^2(2K)$] under\vspace*{1pt} null
hypothesis $H_0$; Stouffer's statistic [$S^{\mathrm{Stouffer}}=\sum\Phi
^{-1}(1-p_k)$, where $\Phi^{-1}(\cdot)$ is the quantile function of a
standard normal distribution] follows\vspace*{1pt} a normal distribution with
variance $K$ [i.e., $S^{\mathrm{Stouffer}}\sim N(0, K)$]; minP statistic
($S^{\mathrm{minP}}=\min\{p_k\}$) follows a Beta distribution with parameters 1
and $K$ [i.e., $S^{\mathrm{minP}}\sim \operatorname{Beta}(1, K)$]; and maxP statistic
($S^{\mathrm{maxP}}=\max\{p_k\}$) follows a Beta distribution with parameters
$K$ and 1 [i.e., $S^{\mathrm{maxP}}\sim \operatorname{Beta}(K, 1)$].

The $\mathrm{HS}_A$ hypothesis setting and maxP method are obviously
too stringent in light of the generally noisy nature of microarray
experiments. When $K$ is large, $\mathrm{HS}_A$ is not robust and
inevitably detects very few genes. Instead of requiring differential
expression in all studies, biologists are more interested in, for
example, ``biomarkers that are differentially expressed in more than
70\% of the combined studies.'' Denote by $\Theta_h= \{\sum_{k=1}^K
I(\theta_{gk}\ne0)=h \}$ the situation that exactly
$h$ out of~$K$ studies are differentially expressed. The new robust
hypothesis setting becomes
\[
\mathrm{HS}_r\dvtx  \Biggl\{H_0\dvtx \bigcap
_k\{\theta_{gk}=0\}\mbox{ versus } H_a^{(r)}\dvtx \bigcup_{h=r}^K
\Theta_h \Biggr\},
\]
where $r=\lceil p\cdot K\rceil$, $\lceil x\rceil$ is the smallest
integer no less than $x$ and $p$ ($0<p\le1$) is the minimal percentage
of studies required to call differential expression (e.g., $p=70\%$).
We note that $\mathrm{HS}_A$ and $\mathrm{HS}_B$ are both special
cases of the extended $\mathrm{HS}_r$ class (i.e., $\mathrm
{HS}_A=\mathrm{HS}_K$ and $\mathrm{HS}_B=\mathrm{HS}_1$), but we
will focus on large $r$ (i.e., $p>50\%$) in this paper and view
$\mathrm{HS}_r$ as a relaxed and robust form of $\mathrm{HS}_A$.

\begin{table}
\tabcolsep=0pt
\tablewidth=250pt
\caption{Four hypothetical genes to compare different meta-analysis methods and to illustrate the motivation of rOP}\label{tsamplegenes}
\begin{tabular*}{\tablewidth}{@{\extracolsep{\fill}}@{}ld{1.4}d{1.3}d{1.4}d{1.4}@{}}
\hline
& \multicolumn{1}{c}{\textbf{Gene A}} & \multicolumn{1}{c}{\textbf{Gene B}} & \multicolumn{1}{c}{\textbf{Gene C}} & \multicolumn{1}{c@{}}{\textbf{Gene D}}\\
\hline
Study 1 & 0.1 & \multicolumn{1}{c}{1e--20\tabnoteref[*]{tt}} & 0.25 & 0.15\\
Study 2 & 0.1 & 0.9 & 0.25 & 0.15\\
Study 3 & 0.1 & 0.9 & 0.25 & 0.15\\
Study 4 & 0.1 & 0.9 & 0.25 & 0.15\\
Study 5 & 0.1 & 0.9 & 0.25 & 0.9\\
Fisher ($\mathrm{HS}_B$) & 0.01\tabnoteref[*]{tt} & \multicolumn{1}{c}{1e--15\tabnoteref[*]{tt}} & 0.18 & 0.12\\
Stouffer ($\mathrm{HS}_B$) & 0.002\tabnoteref[*]{tt} & 0.03\tabnoteref[*]{tt} & 0.07 & 0.10\\
minP ($\mathrm{HS}_B$) & 0.41 & \multicolumn{1}{c}{5e--20\tabnoteref[*]{tt}} & 0.76 & 0.56\\
maxP ($\mathrm{HS}_A$) & \multicolumn{1}{c}{1e--5\tabnoteref[*]{tt}} & 0.59 & 0.001\tabnoteref[*]{tt} & 0.59\\
rOP ($r=4$) ($\mathrm{HS}_r$) & \multicolumn{1}{c}{5e--4\tabnoteref[*]{tt}} & 0.92 & 0.015\tabnoteref[*]{tt} & 0.002\tabnoteref[*]{tt}\\
\hline
\end{tabular*}
\tabnotetext[*]{tt}{$p$-values smaller than 0.05.}
\end{table}

In the literature, maxP has been used for $\mathrm{HS}_A$ and minP has
been used for~$\mathrm{HS}_B$. An intuitive extension of these two
methods for $\mathrm{HS}_r$ is to use the $r$th ordered \mbox{$p$-}value (rOP).
Before introducing the algorithm and properties of rOP, we 
illustrate the motivation of it by the following example.
Suppose we consider four genes in five studies: gene A has marginally
significant $p$-values ($p=0.1$) in all five studies; gene B has a strong
$p$-value in study 1 ($p={}$1e--20) but $p=0.9$ in the other four studies;
gene C is similar to gene~A but has much weaker statistical
significance ($p=0.25$ in all five studies); gene D differs from gene C
in that studies 1--4 have small $p$-values ($p=0.15$) but study 5 has a
large $p$-value ($p=0.9$). Table~\ref{tsamplegenes} shows the resulting
$p$-values from five meta-analysis methods that are derived from
classical parametric inference in Section~\ref{sintro}. Comparing
Fisher and minP in $\mathrm{HS}_B$, minP is sensitive to a study that
has a very small $p$-value (see gene B) while Fisher, as an evidence
aggregation method, is more sensitive when all or most studies are
marginally statistically significant (e.g., gene A). Stouffer behaves
similarly to Fisher except that it is less sensitive to the extremely
small $p$-value in gene B. When we turn our attention to $\mathrm
{HS}_A$, gene C and gene D cannot be detected by all three of the
Fisher, Stouffer and minP methods. Gene C can be detected by both maxP
and rOP as expected ($p=0.001$ and $0.015$, resp.). For gene D,
it cannot be identified by the maxP method ($p=0.59$) but can be
detected by rOP at $r=4$ ($p=0.002$). Gene D gives a good motivating
example that maxP may be too stringent when many studies are combined
and rOP provides additional robustness when one or a small portion of
studies are not statistically significant. In genomic meta-analysis,
genes similar to gene D are common due to the noisy nature of
high-throughput genomic experiments or when a low-quality study is
accidentally included in the meta-analysis. Although the types of
desired markers (under $\mathrm{HS}_A$, $\mathrm{HS}_B$ or $\mathrm
{HS}_r$) depend on the biological goal of a specific application, genes
A, C and D are normally desirable marker candidates that researchers
wish to detect in most situations while gene B is not (unless
study-specific markers are expected as mentioned in Section~\ref
{sintro}). This toy example motivates the development of a robust
order statistic of rOP below.

\subsection{The rOP method}
\label{srop}
Below is the algorithm for rOP when the parameter $r$ is fixed. For a
given gene $g$, without loss of generality, we ignore the subscript $g$
and denote by $S_{r}=p_{(r)}$, where $p_{(r)}$ is the $r$th order
statistic of $p$-values $\{p_{1},p_{2},\ldots,p_{K}\}$. Under the null
hypothesis $H_0$, $S_{r}$ follows a beta distribution with shape
parameters $r$ and $K-r+1$, assuming that the model to generate
$p$-values under the null hypothesis is correctly specified and all
studies are independent. To implement rOP, one may apply this
parametric null distribution to calculate the $p$-values for all genes
and perform a Benjamini--Hochberg (BH) correction\vadjust{\goodbreak} [\citet{benjamini1995controlling}] to control the false discovery rate (FDR)
under the general dependence structure. The Benjamini--Hochberg
procedure can control the FDR at the nominal level or less when the
multiple comparisons are independent or positively dependent. Although
the Benjamini--Yekutieli (BY) procedure can be applied to a more general
dependence structure of the comparisons, it is often too conservative
and unnecessary [\citet{benjamini2001control}], especially in gene
expression analysis where the comparisons are more likely to be
positively dependent and the effect sizes are usually small to moderate
(also see Section~\ref{ssimu} for simulation results). As a result,
we will not consider the BY procedure in this paper. The parametric BH
approach has the advantage of fast computation, but in many situations
the parametric beta null distribution may be violated because the
assumptions to obtain $p$-values from each single study are not met and
the null distributions of $p$-values are not uniformly distributed. When
such violations of assumptions are suspected, we alternatively
recommend a conventional permutation analysis (PA) instead. Class
labels of the samples in each study are randomly permuted and the
entire DE and meta-analysis procedures are followed. The permutation is
repeated for $B$ times ($B=500$ in this paper) to simulate the null
distribution and assess the $p$-values and $q$-values. The permutation
analysis is used for all meta-analysis methods (including rOP, Fisher,
Stouffer, minP and maxP) in this paper unless otherwise stated. 


We note that both minP and maxP are special cases of rOP, but in this
paper we mainly consider properties of rOP as a robust form of maxP
(specifically, $K/2\le r\le K$).

\subsection{Selection of $r$ in an application}
\label{sselectr}
The best selection of $r$ should depend on the biological interests.
Ideally, $r$ is a tuning parameter that is selected by the biologists
based on the biological questions asked and the experimental designs of
the studies. However, in many cases, biologists may not have a strong
prior knowledge for the selection of $r$ and data-driven methods for
estimating $r$ may provide additional guidance in applications. The
purpose of selecting $r<K$ is to tolerate potentially outlying studies
and noises in the data. The noises may come from experimental
limitations (e.g., failure in probe design, erroneous gene annotation
or bias from experimental protocol) or heterogeneous patient cohorts in
different studies. Another extreme case may come from inappropriate
inclusion of a low-quality study into the genomic meta-analysis. Below
we introduce two complementary guidelines to help select $r$ for rOP.
The first method comes from the adjusted number of detected DE genes
and the second is based on pathway association (a.k.a. gene set
analysis), incorporating external biological knowledge.\looseness=-1

\subsubsection{Evaluation based on the number of detected DE genes}
\label{sndet}
\begin{figure}

\includegraphics{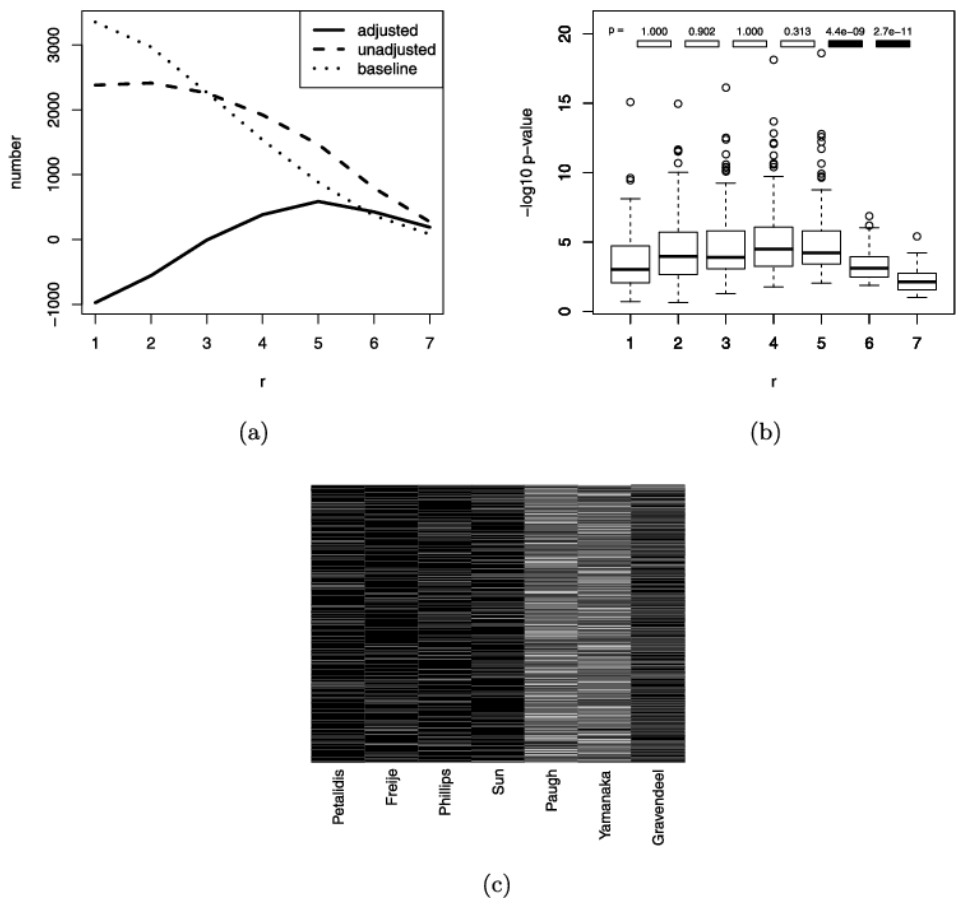}

\caption{Results of brain cancer data set applying rOP.
\textup{(a)}\label{fbraina}~Adjusted and unadjusted number of detected DE genes using different $r$.
\textup{(b)}\label{fbrainb}~Boxplots of $-\log(p)$ for the biological association evaluation. 
$p$-values for Wilcoxon signed-rank tests are shown on the top. Black filled rectangles represent a $p$-value smaller than 0.05.
\textup{(c)}\label{fbrainc}~Heatmap to show effective studies of rOP in each gene. Effective studies are shown in black and noneffective ones are in light gray.}\label{fbrain}
\end{figure}

In the first method, we use a heuristic criterion to find the best $r$
such that the number\vadjust{\goodbreak} of detected DE genes is the largest. The
dashed
line in Figure~\ref{fbrain}(a) shows the number of detected DE genes
using different $r$ in rOP in a brain cancer application. The result
shows a general decreasing trend in the number of detected DE genes
when $r$ increases. However, when we randomly permute the $p$-values
across genes within each study, the detected number of DE genes also
shows a bias toward small $r$'s (dotted line). It shows that a large
number of DE genes can be detected by a small~$r$ (e.g., $r=1$ or 2)
simply by chance. To eliminate this artifact, we apply a detrending
method by subtracting the dotted permuted baseline from the dashed
line. The resulting adjusted number of DE genes (solid line) is then
used to seek the maximum that correspond to the suggested $r$. This
detrend adjustment is similar to what was used in the GAP statistic
[\citet{tibshirani2001estimating}] when estimating the number of
clusters in cluster analysis. In such a scenario, the curve of number
of clusters (on $x$-axis) versus sum of squared within-cluster
dispersions is used to estimate the number of clusters. The curve
always has a decreasing trend even in random data sets and the goal is
usually to find an ``elbow-like'' turning point. The GAP statistic
permutes the data to generate a baseline curve and subtract it from the
observed curve. The problem becomes finding the maximum point in the
detrended curve, a setting very similar to ours.

%

Below we describe the algorithm for the first criterion. Using the
original $K$ studies, the number of DE genes detected by rOP using
different $r$ ($1\le r\le K$) is first calculated as $N_r$ [under the
certain false discovery rate threshold, e.g., \mbox{$\mathrm{FDR}=5\%$}; see
dashed line in Figure~\ref{fbrain}(a)]. We then randomly permute
$p$-values in each study independently and recalculate the number of DE
genes as $N_r^{(b)}$ in the $b$th permutation. The permutation is
repeated for $B$ times ($B=100$ in this paper) and the adjusted number
of detected DE genes is defined as $N'_r=N_r-\sum_{b=1}^B N_r^{(b)}/B$
[see\vspace*{1pt} solid line in Figure~\ref{fbrain}(a)]. In other words, the
adjusted number of DE genes is detrended so that it is purely
contributed by the consistent DE information among studies. The
parameter $r$ is selected so that $N'_r$ is maximized (or we manually
select $r$ as large as possible when $N'_r$ reaches among the largest).

%
\begin{remark}
Note that $N'_r$ could sometimes be negative. This happens mostly when
the signal in a single study is strong and $r$ is small.
However, since we usually apply rOP for relatively large $K$ and $r$,
the negative value is usually not an issue. We also note that, unlike
the GAP statistic, the criterion to choose $r$ with the maximal
adjusted number of detected DE genes is heuristic and has no
theoretical guarantee. In simulations and real applications to be shown
later, this method performs well and provides results consistent with
the second criterion described below.
\end{remark}

\subsubsection{Evaluation based on biological association}
Pathway analysis (a.k.a. gene set analysis) is a statistical\vadjust{\goodbreak} tool to
infer the correlation of differential expression evidence in the data
with pathway knowledge (usually sets of genes with known common
biological function or interactions) from established databases. In
this approach, we hypothesize that the best selection of $r$ will
produce a DE analysis result that generates the strongest statistical
association with ``important'' (i.e., disease-related) pathways. Such
pathways can be provided by biologists or obtained from pathway
databases. However, it is well recognized that our understanding of
biological and disease-related pathways are relatively poor and subject
to change every few years. This is especially true for many complex
diseases, such as cancers, psychiatric disorders and diabetes. In this
case, it is more practical to use computational methods to generate
``pseudo'' disease-related pathways that are further reviewed by
biologists before being utilized to estimate $r$. Below, we develop a
computational procedure for selecting disease-related pathways. We
perform pathway analysis using a large pathway database (e.g., GO, KEGG
or \mbox{BioCarta}) and select pathways that are top ranked by aggregated
committee decision of different $r$ from rOP. The detailed algorithm is
as follows:
\renewcommand{\thestep}{\Roman{step}}
%
\begin{step}
Identification of disease-related pathways (committee decision by
$[K/2]+1\le r\le K$):
\begin{enumerate}[3.]
\item Apply rOP method to combine studies and generate $p$-values for
each gene. Run through different $r$, $[K/2]+1\le r\le K$.
\item For a given pathway $m$, apply Kolmogorov--Smirnov test to compare
the \mbox{$p$-}values of genes in the pathway and those outside the pathway. The
pathway enrichment $p$-values are generated as $p_{r,m}$. Its rank among
all pathways for a given $r$ is calculated as $R_{r,m}=\rank
_m(p_{r,m})$. Small ranks suggest strong pathway enrichment for pathway $m$.
\item The\vspace*{1pt} sums of ranks of different $r$ are calculated as $S_m=\sum
_{r=[K/2]+1}^K R_{r,m}$. The top $U=100$ pathways with the smallest
$S_m$ scores are selected and denoted as~$M$. We treat $M$ as the
``pseudo'' disease-related pathway set.
\end{enumerate}
\end{step}
%
\begin{step}
Sequential testing of improved pathway enrichment significance:
\begin{enumerate}[2.]
\item We perform sequential hypothesis testing that starts from $r'=K$
since conceptually we would like to pick $r$ as large as possible. We
first perform a Wilcoxon signed-rank test to test for difference of
pathway enrichment significance for $r'=K$ and $r'=K-1$. In other
words, we perform a two-sample test on the paired vectors of $(p_{K,m};
m\in M)$ and $(p_{K-1,m};m\in M)$ and record the $p$-value as $\tilde
{p}_{K,K-1}$.
\item If the test is rejected (using the conventional type I error of
0.05), it indicates that reducing from $r=K$ to $r=K-1$ can generate a
DE gene list that produce more significant pathway enrichment in $M$.
We will continue to reduce $r'$ by one (i.e., $r'=K-1$) and repeat the
test between $(p_{r',m};m\in M)$ and $(p_{r'-1,m};m\in M)$. Similarly,
the resulting $p$-values are recorded as $\tilde{p}_{r',r'-1}$. The
procedure is repeated until the test from $r'$ is not rejected. The
final $r'$ is selected for rOP.
\end{enumerate}
\end{step}

\begin{remark}
Note that for simplicity and since this evaluation should be examined
together with the first criterion in Section~\ref{sndet}, we will not
perform $p$-value correction for multiple comparison or sequentially
dependent hypothesis testings here. Practically, we suggest to select
$r$ based on the diagnostic plots of the two criteria simultaneously.
Examples of the selection will be shown in Section~\ref{sapplication}.
\end{remark}

\begin{remark}
We have tested different $U$ in real applications. As can be expected,
the selection of $U$ did not affect the result much. In supplement Figure~7
[\citet{ST2014d}], we show that the ranks for rOP with different selection of
$r$ as well as other methods become stable enough when $U=100$ for all
our applications.
\end{remark}

\subsection{One-sided test modification to avoid discordant effect sizes}\label{srop1side}
Methods combining effect sizes (e.g., random or fixed effects models)
are suitable to combine studies with binary outcome, in which case the
effect sizes are well defined as the standardized mean differences or
odds ratios. Methods combining $p$-values, however, have advantages in
combining studies with nonbinary outcomes (e.g., multi-class,
continuous or censored data), in which case the F-test, simple linear
regression or the Cox proportional hazard model can be used to generate
$p$-values for integration. On the other hand, $p$-value combination
methods usually combine two-sided $p$-values in binary outcome data. A
gene may be found statistically significant with up-regulation in one
study and down-regulation in another study. Such a confusing
discordance, although sometimes a reflection of the biological truth,
is often undesirable in most applications. Therefore, we make a
one-sided test modification to the rOP method similar to the
modification that \citet{owen2009karl} and \citet
{pearson1934new} applied on Fisher's method. The modified rOP statistic
is defined as the minimum of the two rOP statistics combining the
one-sided tests of both tails. Details of this test statistic can be
found in the supplementary material [\citet{ST2014a}].

\section{Applications}\label{sapplication}
We applied rOP as well as other meta-analysis methods to three
microarray meta-analysis applications with different strengths of DE
signal and different degrees of heterogeneity.
Supplement Table 1A--C\vadjust{\goodbreak} [\citet{ST2014c}] list the detailed information on seven brain
cancer studies, nine major depressive disorder (MDD) studies and 16
diabetes studies for meta-analysis. Data were preprocessed and
normalized by standard procedures in each array platform. Affymetrix
data sets were processed by the RMA method and Illumina data sets were
processed by manufacturer's software with quantile normalization for
probe analysis. Probes were matched to the same gene symbols. When
multiple probes (or probe sets) matched to one gene symbol, the probe
that contained the largest variability (i.e., inter-quartile range) was
used to represent the gene. After gene matching and filtering, 5836,
7577 and 6645 genes remained in the brain cancer, MDD and diabetes
data sets, respectively.
The brain cancer studies were collected from the GEO database.
The MDD studies were obtained from Dr. Etienne Sibille's lab. %
A random intercept model adjusted for potential confounders was applied
to each MDD study to obtain $p$-values [\citet{wang2012detecting}].
Preprocessed data of 16 diabetes studies described by \citet
{park2009integration} were obtained from the authors. 
For studies with multiple groups, we followed the procedure of
Park et~al. by taking the minimum $p$-value of all
the pairwise comparisons and adjusted for multiple tests. 
All the pathways used in this paper were downloaded from the Molecular
Signatures Database [MSigDB, \citet{subramanian2005gene}]. Pathway
collections c2, c3 and c5 were used for the $r$ selection purpose.

\subsection{Application of rOP}\label{sapprop}
In all three applications, we demonstrate the estimation of $r$ for rOP
using the two evaluation criteria in Section~\ref{sselectr}. In the
first data set, two important subtypes of brain tumors---anaplastic
astrocytoma (AA) and glioblastoma multiforme (GBM)---were compared in
seven microarray studies. To estimate an adequate $r$ for the rOP
application, we calculated the unadjusted number, the baseline number
from permutation and the adjusted number of detected DE genes using
$1\le r\le7$ under FDR${}=5\%$ [Figure~\ref{fbrain}(a)]. The result showed
a peak at $r=5$. For the second estimation method by pathway analysis,
boxplots of $-\log_{10}(p)$ ($p$-values calculated from association of
DE gene list with top pathways) versus $r$ were plotted [Figure~\ref{fbrain}(b)]. The Wilcoxon signed-rank tests showed that the result from
$r=6$ is significantly more associated with pathways than that from
$r=7$ ($p={}$2.7e--11) and similarly for $r=5$ versus $r=6$ ($p={}$4.4e--9).
Combining the results from Figure~\ref{fbrain}(a)~and~(b),
we decided to choose $r=5$ for this application. Figure~\ref{fbrain}(c)
shows the heatmap of studies effective in rOP (when $r=5$) for each
detected DE gene (a total of 1469 DE genes on the rows and seven
studies on the columns). For example, if $p$-values for the seven studies
are $(0.13, 0.11, 0.03, 0.001, 0.4, 0.7, 0.15)$, the test statistic for
rOP is $S^{\mathrm{rOP}}=0.15$ and the five effective studies that contribute to
rOP are indicated as $(1,1,1,1,0,0,1)$. In the heatmap, effective
studies were indicated by black color and noneffective studies were in
light gray. As shown in Figure~\ref{fbrain}(c), Paugh and Yamanaka were
noneffective studies in almost all detected DE genes, suggesting that
the two studies did not contribute to the meta-analysis and may
potentially be problematic studies. This finding agrees with a recent
MetaQC assessment result using the same seven studies [\citet{kang2012metaqc}]. In our application, AA and GBM patients were
compared in all seven studies. We expected to detect biomarkers that
have consistent fold-change direction across studies and the one-sided
corrected rOP method was more preferable.
Supplement Figure~1 [\citet{ST2014d}] showed plots similar to Figure~\ref{fbrain} for
one-sided corrected rOP. The result similarly concluded that $r=5$ 
was the most suitable choice for this application.


For the second application, nine microarray studies used different
areas of post-mortem brain tissues from MDD patients and control
samples (supplement Table~1B [\citet{ST2014c}]). MDD is a complex genetic disease with
largely unknown disease mechanism and gene regulatory networks. The
postmortem brain tissues usually result in weak signals, compared to
blood or tumor tissues, which makes meta-analysis an appealing
approach. 
In supplement Figure~2(a) [\citet{ST2014d}], 
the maximizer of adjusted DE gene detection was at $r=6$ ($r=7$ or $8$
is also a good choice). For supplement Figure~2(b), 
the statistical significance improved ``from $r=9$ to $r=8$''
($p={}$5.6e--14), ``from $r=8$ to $r=7$'' \mbox{($p={}$8.7e--7)} and ``from $r=7$ to
$r=6$'' ($p=0.045$). We also obtained 98 pathways that were potentially
related to MDD from Dr. Etienne Sibille. As shown in supplement
Figure~2(c), 
the statistical significance improved ``from $r=8$ to $r=7$'' using the
98 expert selected pathways. Combining the results, we decided to
choose $r=7$ (since $r=6$ only provided marginal improvement in both
criteria and we preferred $r$ as large as possible) for the rOP method
in this application. Supplement Figure~2(d) 
showed the heatmap of effective studies in rOP. No obvious problematic
study was observed. The one-sided rOP was also applied (results not
shown); good selection of $r$ appeared to be between 5 and 7.


In the last application, 16 diabetes microarray studies were combined.
These 16 studies were very heterogeneous in terms of the organisms,
tissues and experimental design (supplement Table 1C [\citet{ST2014c}]). 
Supplement Figure~7 [\citet{ST2014d}] showed diagnostic plots to estimate $r$. Although
the number of studies and heterogeneity across data sets were
relatively larger than the previous two examples, we could still
observe similar trends in supplement Figure~7. 
Specifically, for supplement Figure~3(a), 
it was shown that $r=7$--12 detected a higher adjusted number of DE
genes. For pathway analysis, results from $r=12$ were more associated
with the top pathways. As a result, we decided to use $r=12$ in this
application. It was noticeable that the $r$ selection in this diabetes
example was relatively vague compared to the previous examples. 
Supplement Figure~3(c) showed the heatmap of effective studies in rOP.
Two to four studies (s01, s05, s08 and s14) appeared to be candidates
of problematic studies, but the evidence was not as clear as the brain
cancer example in Figure~\ref{fbrain}(c). It should be noted that the
results of supplement Figure~3 
used the beta null distribution inference and Benjamini--Hochberg
correction. 
Permutation analysis generated a relatively unstable result (supplement
Figure~4), although it suggested a similar selection of $r$. This was
possibly due to the unusual \textit{ad hoc} DE analysis from minimum
$p$-values of all possible pairs of comparisons [procedures that were
used in the original paper \citet{park2009integration}].

Next, we explored the robustness of rOP by mixing a randomly chosen MDD
study into seven brain cancer studies as an outlier. The results in
supplement Figure~5 [\citet{ST2014d}] 
showed that $r=5$ or $6$ may be a good choice [supplement Figure~5(a)~and~(b)].
We used $r=6$ in rOP for this application. 
Supplement Figure~5(c) interestingly showed that the mixed MDD study,
together with the Paugh and Yamanaka studies, was a potentially
problematic study in the rOP meta-analysis. This result verified our
intuition that rOP is robust to outlying studies and the $p$-values of
the outlying studies minimally contribute to the rOP \mbox{statistic.}

\subsection{Comparison of rOP with other meta-analysis methods}
$\!\!$We performed rOP using $r$ determined from Section~\ref{sapprop} in
four applications (brain cancer, MDD, diabetes and brain cancer $+$\,1
random MDD) and compared to Fisher's method, Stouffer's method, minP,
maxP and vote counting. The vote counting method will be discussed in
greater detail in Section~\ref{svc}. Two quantitative measures were
used to compare the methods. The first measure compared the number of
detected DE genes from each method as a surrogate of sensitivity
(although the true list of DE genes is unknown and sensitivity cannot
be calculated). The second approach was by pathway analysis, very
similar to the method we introduced to select parameter~$r$. However,
in order to avoid bias in top pathway selection, single study analysis
results were used as the committee to select disease-related pathways.
KEGG, BioCarta, Reactome and GO pathways were used in the pathway
analysis. The Wilcoxon signed-rank test was then used to test if two
methods detected DE genes with differential association with
disease-related pathways.

%
\begin{table}
\tabcolsep=3pt
\caption{Number of DE genes detected by different methods under FDR${}=5\%$}\label{tndet}
\begin{tabular*}{\tablewidth}{@{\extracolsep{\fill}}@{}lccccd{4.0}d{3.0}d{3.0}@{}}
\hline
& \multicolumn{2}{c}{\textbf{rOP}}\\[-6pt]
& \multicolumn{2}{c}{\hrulefill}\\
& \textbf{Two-sided} & \textbf{One-sided} & \textbf{Fisher} & \textbf{Stouffer} & \multicolumn{1}{c}{\textbf{minP}} & \multicolumn{1}{c}{\textbf{maxP}} & \multicolumn{1}{c@{}}{\textbf{VC}}\\
\hline
Brain cancer & 1469 ($r=5$) & 1625 ($r=5$) & 2918 & 2449 & 2380 & 273 & 328\\
& \multicolumn{2}{c}{Overlap${}={}$1139}\\[3pt]
MDD & \phantom{0}617 ($r=7$) & \phantom{00}86 ($r=7$) & 1124 & 1423 & 0 & 310 & 0\\
& \multicolumn{2}{c}{Overlap${}={}$48}\\[3pt]
Diabetes & \phantom{10}636 ($r=12$) & Not applicable & 1698 & 1492 & 1 & 85 & 0\\
Brain${}+{}$1 MDD & \phantom{0}751 ($r=6$) & Not applicable & 2081 & 1773 & 1648 & 132 & 64\\
\hline
\end{tabular*}
\end{table}

%
\begin{figure}

\includegraphics{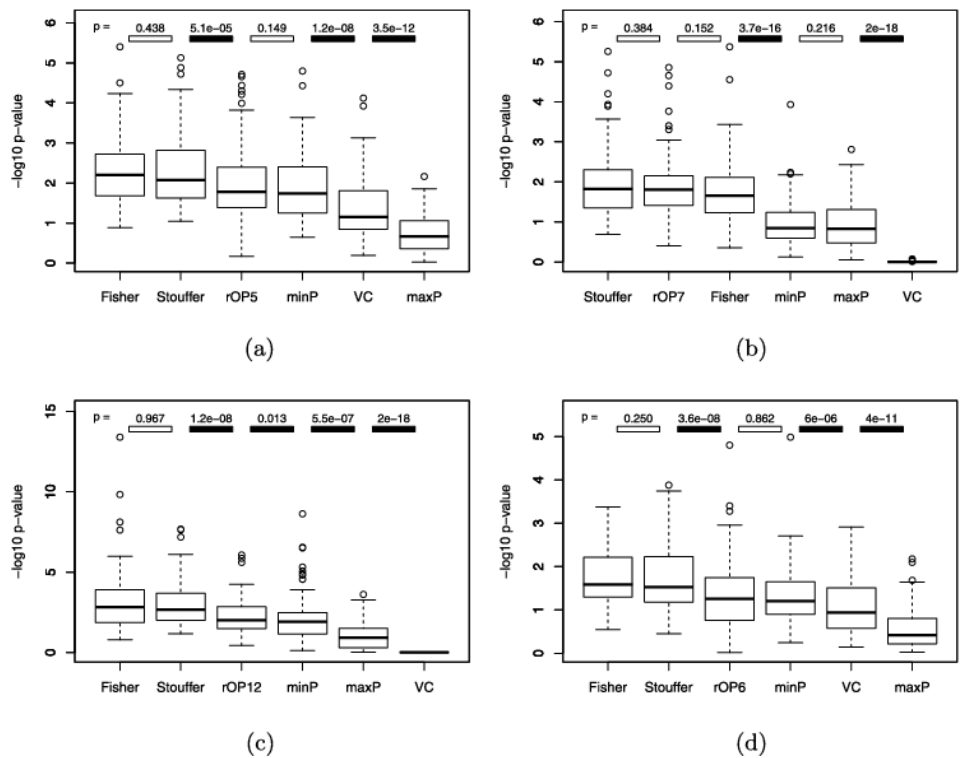}

\caption{Comparison of different meta-analysis methods using pathway analysis.
\textup{(a)}\label{fbraincomp}~Brain cancer.
\textup{(b)}\label{fmddcomp}~MDD.
\textup{(c)}\label{fdiabcomp}~Diabetes.
\textup{(d)}\label{fartcomp}~Brain cancer and 1 random MDD.}\label{fcomp}
\end{figure}

Table~\ref{tndet} showed the number of detected DE genes under FDR${}=5\%$.
We can immediately observe that Fisher and Stouffer generally
detected many more biomarkers because they targeted on $\mathrm{HS}_B$
(genes differentially expressed in one or more studies). Although minP
also targeted on $\mathrm{HS}_B$, it sometimes detected extremely
small numbers of DE genes in weak-signal data such as the MDD and
diabetes examples. This is reasonable because minP has very weak power
to detect consistent but weak signals across studies [e.g.,
$p$-values${}=(0.1, 0.1, \ldots, 0.1)$]. The stringent maxP method detected
few numbers of DE genes in general. Vote counting detected very few
genes especially when the effect sizes were moderate (in~the MDD and
diabetes examples). rOP detected more DE genes than maxP because of its
relaxed $\mathrm{HS}_r$ hypothesis setting. It identified about
50--65\% fewer DE genes than Fisher's and Stouffer's methods, but
guaranteed that the genes detected were differentially expressed in the
majority of the studies. We also performed the one-sided corrected rOP
for comparison. This method detected similar numbers of DE genes
compared to two-sided rOP, and the majority of detected DE genes in
two-sided and one-sided rOP were overlapped in the brain cancer
example. The result showed that almost all DE genes detected by
two-sided rOP had a consistent fold-change direction across studies. In
MDD, the one-sided rOP detected much fewer genes than the two-sided
method. This implied that many genes related to MDD acted differently
in different brain regions and in different cohorts.

Figure~\ref{fcomp} showed the results of biological association from
pathway analysis that were similarly shown in Figure~\ref{fbrain}(b). The
result showed that the DE gene lists generated by Fisher and Stouffer
were more associated with biological pathways. The rOP method generally
performed better than maxP and minP and had similar biological
association performance to Fisher's and Stouffer's methods.

\section{Statistical properties of rOP}
\label{sstat}
\subsection{Power calculation of rOP and asymptotic properties}
\label{spowcalc}
When $K$ studies are combined, suppose $r_0$ of the $K$ studies have
equal nonzero effect sizes and the rest of the ($K-r_0$) studies have
zero effect sizes. That is,
\begin{eqnarray*}
H_0\dvtx \theta_1&=&\cdots=\theta_K=0,
\\
H_a\dvtx \theta_1&=&\cdots=\theta_{r_0}=\theta
\ne0,\qquad \theta_{r_0+1}=\cdots=\theta_K=0.
\end{eqnarray*}

For a single study, the power function given effect size $\theta$ is
known as $\Pr(p_i\le\alpha_0|\theta)$. We will derive the
statistical power of rOP under this simplified hypothesis setting when
$r_0$ and $r$ for rOP are given. Under $H_0$, the rejection threshold
for the rOP statistic is $\beta=B_\alpha(r,K-r+1)$ (the $\alpha$
quantile of a beta distribution with shape parameters $r$ and $K-r+1$),
where the significance level of the meta-analysis is set at $\alpha$.
The power of rejection threshold $\beta$ under $H_a$ is $\Pr
(p_{(r)}\le\beta|H_a )=\Pr(\sum_{i=1}^K I(p_i\le\beta
)\ge r|H_a )$. By definition, $\Pr(p_i\le\beta|\theta
_i=0)=\beta$ and we further denote $\beta'=\Pr(p_i\le\beta|\theta
_i=\theta)$. The power calculation of interest is equivalent to
finding the probabilities of having at least $r$ successes in $K$
independent Bernoulli trials, among which $r_0$ have success
probabilities $\beta'$, and $K-r_0$ have success probabilities $\beta$:
\begin{eqnarray*}
\Pr(p_{(r)}\le\beta|H_a )&=&\sum
_{i=r}^K\sum_{j=\max
(0,i-K+r_0)}^{\min(i,r_0)}\pmatrix{r_0
\cr j}\beta'^j\bigl(1-\beta'
\bigr)^{r_0-j}
\\
&&\hspace*{84pt}{}\times
\pmatrix{K-r_0 \cr i-j}\beta^{i-j}(1-\beta)^{K-r_0-i+j}.
\end{eqnarray*}

\begin{remark}
We note that the assumption of $r_0$ equal nonzero effect sizes can be
relaxed. When the nonzero effects are not equal, the power calculation
can be done in polynomial time using dynamic programming.
\end{remark}

\begin{figure}

\includegraphics{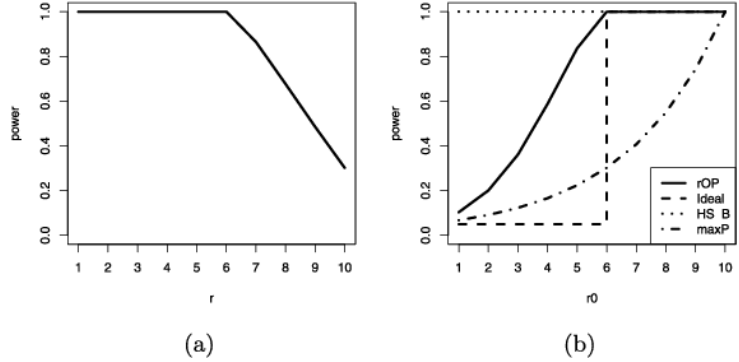}

\caption{Power of rOP method when $N_k\rightarrow\infty$, $K=10$.
\textup{(a)}\label{fpowr}~$r_0=6$, $r=1$--10.
\textup{(b)}\label{fpowr0}~$r=6$, $r_0=0$--10.}\label{fpow}
\end{figure}

Below we demonstrate some asymptotic properties of rOP.
%
\begin{theorem}
\label{tpow1}
Assume $r_0$ is fixed. When the effect size $\theta$ and $K$ are fixed
and the sample size of study $k$ $N_k\rightarrow\infty$, $\Pr
(p_{(r)}\le\beta|H_a )\rightarrow1$ if $r\le r_0$. When
$r>r_0$, $\Pr(p_{(r)}\le\beta|H_a )\rightarrow c(r)<1$
and $c(r)$ is a decreasing function in $r$.
\end{theorem}
\begin{pf}
When $N_k\rightarrow\infty$, $\beta'\rightarrow1$. The theorem
easily follows from the power calculation formulae.
\end{pf}
Theorem \ref{tpow1} states that, asymptotically, if the parameter $r$
in rOP is specified less or equal to the true $r_0$, the statistical
power converges to 1 as intuitively expected. When specifying $r$
greater than $r_0$, the statistical power is weakened with increasing
$r$. Particularly, maxP will have weak power. In contrast to Theorem~\ref{tpow1}, for methods designed for $\mathrm{HS}_B$ (e.g.,
Fisher's method, Stouffer's method and minP), the power always
converges to 1 if $N_k\rightarrow\infty$ and $r_0>0$. Figure~\ref{fpow}(a) shows the power curve of rOP for different $r$ when $K=10$,
$r_0=6$ and $N_k\rightarrow\infty$.

\begin{lemma}
\label{lpow1}
Assume the parameter $r$ used in rOP is fixed. When the effect size
$\theta$ and $K$ are fixed and the sample sizes $N_k\rightarrow\infty
$, $\Pr(p_{(r)}\le\beta|H_a )\rightarrow1$ if $r_0\ge
r$. When $r_0<r$, $\Pr(p_{(r)}\le\beta|H_a )\rightarrow
c(r_0)<1$ and $c(r_0)$ is an increasing function in $r_0$.
\end{lemma}

Lemma \ref{lpow1} takes a different angle from Theorem \ref{tpow1}.
When the parameter $r$ used in rOP is fixed, it asymptotically has
perfect power to detect all genes that are differentially expressed in
$r$ or more studies. It then does not have strong power to detect genes
that are differentially expressed in less than $r$ studies. Figure~\ref{fpow}(b) shows a power curve of rOP for $K=10$, $r=6$ and
$N_k\rightarrow\infty$ (solid line). We note that the dashed line
[$f(r)=0$ when $0\le r_0<6$ and $f(r)=1$ when $6\le r_0\le10$] is the
ideal power curve for $\mathrm{HS}_r$ (i.e., it detects all genes that
are differentially expressed in $r$ or more studies but does not detect
any genes that are differentially expressed in less than $r$ studies).
Methods like Fisher, Stouffer and minP target on $\mathrm{HS}_B$ and
their power is always 1 asymptotically when $r_0>0$. The maxP method
has perfect asymptotic power when $r_0=K=10$ but has relatively weak
power when $r_0<K$. The rOP method lies between maxP and the methods
designed for~$\mathrm{HS}_B$. The power of rOP for $r_0\ge6$
converges to 1, and for $r_0\le5$, the power is always smaller than 1
as the sample sizes in single studies go to infinity. Although the
asymptotic powers of rOP for $r_0=4$ and $r_0=5$ are not too small, we
are less concerned about these genes because they are still very likely
to be important biomarkers.

\subsection{Power comparison in simulated studies}
\label{ssimu}
To evaluate the performance of rOP in the genomic setting, we simulated
a data set using the following procedure.

\setcounter{step}{0}
%
\begin{step}
Sample 200 gene clusters, with 20 genes in each and another 6000 genes
that do not belong to any cluster. Denote $C_g\in\{0, 1, 2, \ldots,
200\}$ as the cluster membership of gene $g$, where $C_g=0$ means that
gene $g$ is not in a gene cluster.
\end{step}
%
\begin{step}
Sample the covariance matrix $\Sigma_{ck}$ for genes in cluster $c$
and in study~$k$, where $1\le c \le200$ and $1\le k\le10$. First,
sample $\Sigma'_{ck}\sim W^{-1}(\Psi, 60)$, where $\Psi
=0.5I_{20\times20}+0.5J_{20\times20}$, $W^{-1}$ denotes the inverse
Wishart distribution, $I$ is the identity matrix and $J$ is the matrix
with all the elements equal 1. Then $\Sigma_{ck}$ is calculated by
standardizing $\Sigma'_{ck}$ such that the diagonal elements are all
1's. 
\end{step}
%
\begin{step}
Denote $g_{c1},\ldots,g_{c20}$ as the indices for the 20 genes in
cluster $c$, that is, $C_{g_{cj}}=c$, where $1\le c\le200$ and $1\le
j\le20$. Assuming the effect sizes are all zeros, sample gene
expression levels of genes in cluster $c$ for sample $n$ as
$(X'_{g_{c1}nk},\ldots,X'_{g_{c20}nk})^T\sim \mathit{MVN}(0, \Sigma_{ck})$,
where\vspace*{2pt} $1\le n\le100$ and $1\le k \le10$, and sample expression level
for gene $g$ which is not in a cluster (i.e., $C_g=0$) for sample $n$
as $X'_{gnk}\sim N(0, 1)$, where $1\le n\le100$ and $1\le k\le10$.
\end{step}
%
\begin{step}
Sample the true number of studies that gene $g$ is DE, $t_g$, from a~discrete uniform distribution that takes values on $1,2,\ldots,10$,
for $1\le g\le 1000$; and set $t_g=0$ for $1001\le g\le{}$10,000.
\end{step}
%
\begin{step}
Sample $\delta_{gk}$, which indicates whether gene $g$ is DE in study
$k$, from a discrete uniform distribution that takes values on 0 or 1
and with the constraint that $\sum_k \delta_{gk}=t_g$, where $1\le
g\le1000$ and $1\le k\le10$. For $1001\le g\le{}$10,000 and $1\le
k\le10$, set $\delta_{gk}=0$.
\end{step}
%
\begin{step}
Sample the effect size $\mu_{gk}$ uniformly from $[-1,-0.5]\cup[0.5,1]$.
For control samples, set the expression levels as $X_{gnk}=X'_{gnk}$;
for case samples, set the expression levels as
$Y_{gnk}=X'_{g(n+50)k}+\mu_{gk}\cdot\delta_{gk}$, for $1\le g\le{}$10,000, $1\le n\le50$ and $1\le k\le10$.
\end{step}

\begin{table}
\tabcolsep=0pt
\caption{Mean FDRs for different methods in $\mathrm{HS}_r$ with
$r=6$ by simulation analysis with correlated genes. The standard
deviations of the FDRs in using 100 simulations are shown in the parentheses}\label{tfdr12}
\begin{tabular*}{\tablewidth}{@{\extracolsep{\fill}}lccc@{}}
\hline
& $\mathbf{FDR_1}$ & $\mathbf{FDR_2}$ & \textbf{\# of detected genes}\\
\hline
rOP ($r=6$, PA) & 0.0439 ($\pm$\,0.0106) & 0.1818 ($\pm$\,0.0179) & 620.16\\
rOP ($r=6$, BH) & 0.0472 ($\pm$\,0.0094) & 0.2029 ($\pm$\,0.0184) & 617.53\\
rOP ($r=6$, BY) & 0.0043 ($\pm$\,0.0031) & 0.1044 ($\pm$\,0.0139) & 539.85\\
Fisher & 0.0441 ($\pm$\,0.0090) & 0.4186 ($\pm$\,0.0212) & 934.91\\
Stouffer & 0.0440 ($\pm$\,0.0089) & 0.3623 ($\pm$\,0.0217) & 858.86\\
minP & 0.0466 ($\pm$\,0.0103) & 0.4567 ($\pm$\,0.0207) & 958.26\\
maxP & 0.0459 ($\pm$\,0.0199) & 0.0729 ($\pm$\,0.0251) & 201.02\\
Vote counting & 0.0000 ($\pm$\,0.0000) & 0.0003 ($\pm$\,0.0016) & 234.43\\
\hline
\end{tabular*}
\end{table}

\begin{figure}[t]

\includegraphics{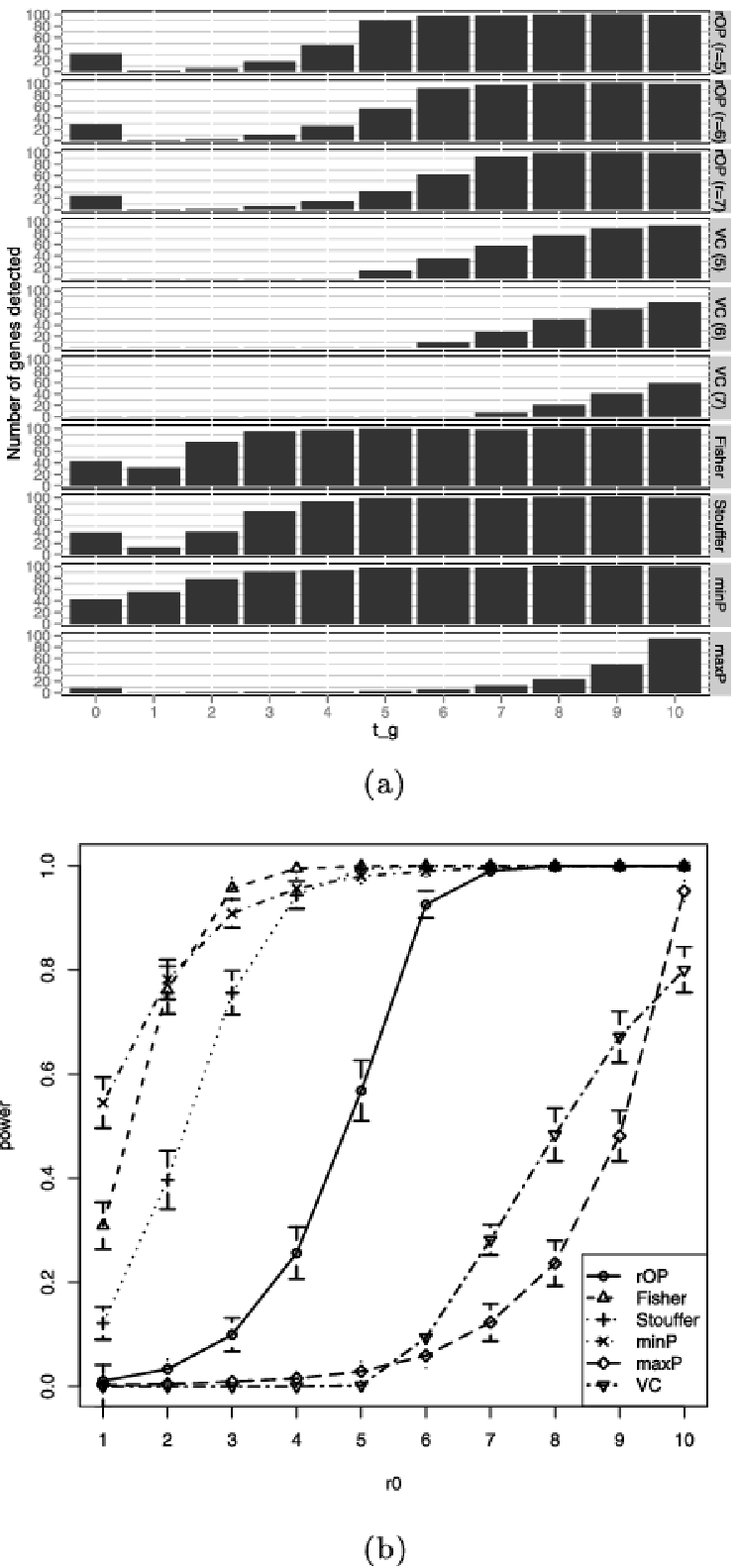}

\caption{Simulation results for rOP and other methods with correlated genes.
\textup{(a)}\label{fsimundet}~Number of genes detected by difference methods. The detected genes are binned according to their $t_g$'s.
\textup{(b)}\label{fsimupow}~Power of different methods for genes with $r_0$ nonzero effect sizes.}\vspace*{-2pt}\label{fsimures}
\end{figure}

In the simulated data set, 10 studies with 10,000 genes were simulated.
Within each study, there were 50 cases and 50 controls. The first 1000
genes were DE in~1 to 10 studies with equal probabilities; and the rest
of the 9000 genes were DE in none of the studies. We denoted $t_g$ as
the true number of studies where gene $g$ was~DE. To mimic the gene
dependencies in a real gene expression data set, within the 10,000
genes, we drew 200 gene clusters with 20 genes in each. We sampled the
data such that the genes within the same cluster were correlated. The
correlation matrices for different studies and different gene clusters
were sampled from an inverse Wishart distribution. Suppose the goal of
the meta-analysis was to obtain biomarkers differentially expressed in
at least 60\% (6 out of 10) of the studies (i.e., $\mathrm{HS}_r$ with
$r=6$). We performed two sample $t$-tests in each study and combined the
$p$-values using rOP with $r=6$. $\mathrm{FDR}\le5\%$ was controlled
using the permutation analysis. To compare rOP with other methods in
the $\mathrm{HS}_r$ setting, we defined two FDR criteria as follows.
Note that $\mathrm{FDR}_1$ targets on $H_0\dvtx t_g=0$ and $\mathrm
{FDR}_2$ targets on $H_0\dvtx t_g<r$:
\begin{eqnarray*}
\mathrm{FDR}_1&=&\frac{\sum_gI(t_g=0 \mbox{ and gene $g$ is
detected})}{\#\{\mbox{genes detected}\}},
\\
\mathrm{FDR}_2&=&\frac{\sum_gI(t_g<r \mbox{ and gene $g$ is
detected})}{\#\{\mbox{genes detected}\}}.
\end{eqnarray*}
Table~\ref{tfdr12} listed the average $\mathrm{FDR}_1$ and $\mathrm
{FDR}_2$ for different methods calculated using 100 simulations. We can
see that although $\mathrm{FDR}_1$ was well controlled, all the
methods were anti-conservative in terms of $\mathrm{FDR}_2$, since the
inference of the five methods was based on $H_0\dvtx t_g=0$ while genes with
$1\le t_g\le5$ existed and were calculated toward $\mathrm{FDR}_2$.
To compare different FDR control methods, we also included the results
of the Benjamini--Hochberg and Benjamini--Yekutieli procedures. According
to the simulation, the Benjamini--Hochberg procedure controlled FDR
similarly to the permutation test. The Benjamini--Yekutieli procedure,
on the other hand, was too conservative that the $\mathrm{FDR}_1$ was
controlled at about $1/10$ of the nominal FDR level. Figure~\ref
{fsimures} showed the number of detected DE genes and the statistical
power of different methods for genes with $t_g$ from 1 to 10.
From Figure~\ref{fsimures}(a), we noticed that Fisher, Stouffer and
minP methods detected many genes with $1\le t_g\le5$, which violated
our targeted $\mathrm{HS}_r$ with $r=6$. MaxP detected very few genes
and missed many targeted markers with $6\le t_g\le9$. Only rOP
generated the result most compatible with $\mathrm{HS}_r$ ($r=6$).
Most genes with $6\le t_g\le10$ were detected. The high $\mathrm
{FDR}_2=18.2\%$ mostly came from genes with $4\le t_g\le5$, genes that
were very likely important markers and were minor mistakes. Vote
counting detected genes with $t_g\ge6$ but was less powerful. The
relationship of vote counting and rOP will be further discussed in
Section~\ref{svc}. We also performed rOP ($r=5$) and rOP ($r=7$) to
compare the robustness of slightly different selections of $r$. Among
the 620.16 DE genes (averaged over 100 simulations) detected by rOP
($r=6$), 594.15 (95.8\%) of them were also detected by rOP ($r=5$) and
516.28 (83.3\%) of them were also detected by rOP ($r=7$).
The result of Figure~\ref{fsimures}(b) was consistent with the
theoretical power calculation as shown in Figure~\ref{fpow}(b).

We also performed the simulation without correlated genes. The results
were shown in the supplement Table~2 [\citet{ST2014c}] and supplement Figure~6 [\citet{ST2014d}]. We
noticed that the FDRs were controlled well in both correlated and
uncorrelated cases. However, the standard deviations of FDRs with
correlated genes were higher than the FDRs with only independent genes,
which indicated some instability of the FDR control with correlated
genes reported by \citet{qiu2006some}.

\subsection{Connection with vote counting}\label{svc}
Vote counting has been used in many meta-analysis applications due to
its simplicity, while it has been criticized as being problematic and
statistically inefficient. \citet{hedges1980vote} showed that the
power of vote counting converges to 0 when many studies of moderate
effect sizes are combined (see supplement Theorem 1 [\citet{ST2014b}]). We, however,
surprisingly found that rOP has a close connection with vote counting,
and rOP can be viewed as a generalized vote counting with better
statistical properties. There are many variations of vote counting in
the literature. One popular approach is to count the number of studies
that have $p$-values smaller than a prespecified threshold, $\alpha$. We
define this quantity as
%
\begin{equation}
\label{ervc} r=f(\alpha)=\sum_{k=1}^KI
\{p_k<\alpha\}
\end{equation}
and define its related proportion as $\pi=E(r)/K$. The test hypothesis is
\[
\cases{ H_0\dvtx \pi=\pi_0,
\vspace*{2pt}\cr
H_a\dvtx \pi>
\pi_0,}
\]
where $\pi_0=0.5$ is often used in the applications. Under the null
hypothesis, $r\sim \mathit{BIN}(K, \alpha)$ and $\pi=\alpha$, so the
rejection region can be established. In the vote counting procedure,
$\alpha$ and $\pi_0$ are two preset parameters and the inference is
made on the test statistic $r$.\vadjust{\goodbreak}

In the rOP method, we view equation (\ref{ervc}) from another
direction. We can easily show that if we solve equation (\ref{ervc})
to obtain $\alpha=f^{-1}(r)$, the solution will be $\alpha\in
[p_{(r)},p_{(r+1)} )$, and one may choose $\alpha=p_{(r)}$ as
the solution. In other words, rOP presets $r$ as a given parameter, and
the inference is based on the test statistic \mbox{$\alpha=p_{(r)}$}.





It is widely criticized that vote counting is powerless because when
the effect sizes are moderate and the power of single studies is lower
than $\pi_0$, as $K$ increases, the percentage of significant studies
will converge to the single study power. However, in the rOP method,
because the $r$th quantile is used, tests of the top $r$ studies are
combined, which helps the rejection probability of rOP achieve 1 as
$K\rightarrow\infty$. It should be noted that the major difference
between rOP and vote counting is that the test statistic $\alpha
=p_{(r)}$ in rOP increases as $K$ and $r=K\cdot c$ increase, which
keeps information of the $r$ smallest $p$-values. On the contrary, for
vote counting, $\alpha$ is often chosen small and fixed when $K$
increases. 
In supplement Theorem 1 [\citet{ST2014b}], the power of vote counting converges to 0 as
$K\rightarrow\infty$, while the power of rOP converges to 1
asymptotically as proved in supplement Theorem~2 [\citet{ST2014b}].

\section{Conclusion}\label{sconclusion}
In this paper we proposed a general class of order statistics of
$p$-values, called $r$th ordered $p$-value (rOP), for genomic
meta-analysis. This family of statistics included the traditional
maximum $p$-value (maxP) and minimum $p$-value (minP) statistics that
target on DE genes in ``all studies'' ($\mathrm{HS}_A$) or ``one or
more studies'' ($\mathrm{HS}_B$). We extended $\mathrm{HS}_A$ to a
robust form that detected DE genes ``in the majority of studies''
($\mathrm{HS}_r$) and developed the rOP method for this purpose. The
new robust hypothesis setting has an intuitive interpretation and is
more adequate in genomic applications where unexpected noise is common
in the data. We developed the algorithm of rOP for microarray
meta-analysis and proposed two methods to estimate $r$ in real
applications. Under ``two-class'' comparisons, we proposed a one-sided
corrected form of rOP to avoid detection of discordant expression
change across studies (i.e., significant up-regulation in some studies
but down-regulation in other studies). Finally, we performed power
analysis and examined asymptotic properties of rOP to demonstrate
appropriateness of rOP for $\mathrm{HS}_r$ over existing methods such
as Fisher, Stouffer, minP and maxP. We further showed a surprising
connection between vote counting and rOP that rOP can be viewed as a
generalized vote counting with better statistical property.
Applications of rOP to three examples of brain cancer, major depressive
disorder (MDD) and diabetes showed better performance of rOP over maxP
in terms of detection power (number of detected markers) and biological
association by pathway analysis.

There are two major limitations of rOP. First, rOP is for $\mathrm
{HS}_r$, but the null and alternative hypotheses are not
complementary\vadjust{\goodbreak}
(see Section~\ref{shs}). Thus, it has weaker ability to exclude
markers that are differentially expressed in ``less than $r$'' studies
since the null of $\mathrm{HS}_r$ is ``differentially expressed in
none of the studies.'' One solution to improve the anti-conservative
inference (which is also our future work) is by Bayesian modeling of
$p$-values with a family of beta distributions [\citet{erickson2009meta}]. Second, selection of $r$ may not always be
conclusive from the two methods we proposed; the external pathway
information may especially be prone to errors and may not be
informative to the data. But since choosing slightly different $r$
usually gives similar results, this is not a severe problem in most
applications. We have tested a different approach by adaptively
choosing the best gene-specific $r$ that generates the best $p$-value.
The result is, however, not stable and the gene-specific parameter $r$
is hard to interpret in applications.

Although many meta-analysis methods have been proposed and applied to
microarray applications, it is still not clear which method enjoys
better performance under what condition. The selection of an adequate
(or best) method heavily depends on the biological goal (as illustrated
by the hypothesis settings in this paper) and the data structure. In
this paper, we stated a robust hypothesis setting ($\mathrm{HS}_r$)
that is commonly targeted in biological applications (i.e., identify
markers statistically significant in the majority of studies) and
developed an order statistic method (rOP) as a solution. The three
applications covered ``cleaner'' data (brain cancer) to ``noisier''
data (complex genetics in MDD and diabetes), and rOP performed well in
all three examples. We expect that the robust hypothesis setting and
the order statistic methodology will find many more applications in
genomic research and traditional univariate meta-analysis in the future.

For multiple comparison control, we propose to either apply the
parametric beta null distribution to assess the $p$-value and perform the
Benjamini--Hochberg (BH) procedure for $p$-value adjustment or conduct a
conventional permutation analysis by permuting class labels in each
study. The former approach is easy to implement, and the latter
approach better preserves the gene correlation structure in the
inference. Instead of the BH procedure, we also tested the
Benjamini--Yekutieli (BY) procedure which is applicable to the general
dependence structure but found that it is overly conservative for
genomic applications. The problem of FDR control under general
high-dimensional dependence structures is beyond the scope of this
paper but is critical in applications and deserves future research.

Implementation of rOP is available in the ``MetaDE'' package in R
together with over 12 microarray meta-analysis methods in the
package.\break 
MetaDE has been integrated with other quality control methods
[``MetaQC'' package, \citet{kang2012metaqc}] and pathway enrichment
analysis methods [``MetaPath'' package, \citet{shen2010meta}]. The
future plan is to integrate the three packages with other genomic
meta-analysis tools into a ``MetaOmics'' software suite [\citet{wang2012r}].


\section*{Acknowledgments}
The authors would like to thank Etienne Sibille and Peter Park for
providing the organized major depressive disorder and diabetes data. We
would also like to thank the anonymous Associate Editor and Editor
Karen Kafadar for many suggestions and critiques to improve the paper.


%
\begin{supplement}[id=supptext]
\sname{Supplement Text}
\stitle{Supplement Text\\}
\slink[doi]{10.1214/13-AOAS683SUPPA} 
\sdatatype{.pdf}
\sfilename{aoas683\_suppa.pdf}
\sdescription{Details of one-sided test modification to avoid discordant effect sizes.}
\end{supplement}
\begin{supplement}[id=suppthe]
\sname{Supplement Theorems}
\stitle{Supplement Theorems 1 and 2\\}
\slink[doi]{10.1214/13-AOAS683SUPPB} 
\sdatatype{.pdf}
\sfilename{aoas683\_suppb.pdf}
\sdescription{Theorem 1---Asymptotic property of vote counting as
$K\rightarrow\infty$. Theorem 2---Asymptotic property of rOP as $K\rightarrow\infty$.}
\end{supplement}
\begin{supplement}[id=supptab]
\sname{Supplement Tables}
\stitle{Supplement Tables~1 and 2\\}
\slink[doi,text={10.1214/13-\break AOAS683SUPPC}]{10.1214/13-AOAS683SUPPC} 
\sdatatype{.pdf}
\sfilename{aoas683\_suppc.pdf}
\sdescription{Table~1---Detail information of combined data sets.
Table~2---FDRs for simulation analysis without correlated genes.}
\end{supplement}
\begin{supplement}[id=suppfig]
\sname{Supplement Figures}
\stitle{Supplement Figures~1 to 7\\}
\slink[doi,text={10.1214/13-\break AOAS683SUPPD}]{10.1214/13-AOAS683SUPPD} 
\sdatatype{.pdf}
\sfilename{aoas683\_suppd.pdf}
\sdescription{Figure~1---Results of brain cancer data set using
one-sided corrected rOP.
Figure~2---Results of MDD data set.
Figure~3---Results of diabetes data set.
Figure~4---Permutation results of diabetes data set.
Figure~5---Results of brain cancer and 1 random MDD data set.
Figure~6---Simulation results without correlated genes.
Figure~7---Mean rank of different methods for the top $U$ pathways.}
\end{supplement}


\printaddresses

\end{document}